\documentstyle[seceq,supplement,epsf]{ptptex}

\title{
Study of QCD thermodynamics at finite density \\ by Taylor expansion
}

\author{
Shinji~{\sc Ejiri},$^{1,}$\footnote{Presented by S. Ejiri.} 
Chris~R.~{\sc Allton},$^{2}$ 
Simon~J.~{\sc Hands},$^{2}$ 
Olaf~{\sc Kaczmarek},$^{1}$ 
Frithjof~{\sc Karsch},$^{1}$ 
Edwin~{\sc Laermann},$^{1}$ and 
Christian~{\sc Schmidt}$^{3}$
}

\inst{
$^{1}$ Fakult\"{a}t f\"{u}r Physik, 
Universit\"{a}t Bielefeld, D-33615 Bielefeld, Germany
\\
$^{2}$ Department of Physics, University of Wales Swansea, 
Singleton Park, Swansea, SA2 8PP, U.K.
\\
$^{3}$ Fakult\"{a}t f\"{u}r Physik, Universit\"{a}t Wuppertal, 
D-42097 Wuppertal, Germany
}

\recdate{
\today
}

\abst{
We discuss the phase structure and 
the equation of state for QCD at non-zero temperature and density. 
Derivatives of $\ln Z$ with respect to 
quark chemical potential $\mu_q$ up to fourth order are calculated 
for 2-flavor QCD, 
enabling estimates of the pressure, quark number density and 
associated susceptibilities as functions of $\mu_q$ via a Taylor 
series expansion. 
Also, the phase transition line for 2 and 3-flavor QCD 
and the critical endpoint in the $(T, \mu_q)$ plane are investigated 
in the low density regime. 
}

\begin{document}

\maketitle

\section{Introduction}
\label{sec:intro}

We would like to report our recent study of 
QCD thermodynamics with a small but non-zero quark chemical potential 
$\mu_q$ by numerical simulations. 
The interesting regime for heavy-ion collision experiments is rather 
low density regime, e.g. $\mu_q/T_c \sim 0.1$ for RHIC and 
$\mu_q/T_c \sim 0.5$ for SPS. 
The phase transition for 2-flavor QCD is known to be crossover 
at $\mu_q=0$ and expected to become a first order phase 
transition at a critical endpoint.
To find the endpoint is also an interesting topic in the low density 
regime, and it might be possible to detect the endpoint experimentally 
via event-by-event fluctuations in heavy-ion collisions.

We discuss the equation of state at non-zero 
baryon number density in Sec.~\ref{sec:pressure}\cite{us03}. 
The study of the equation of state gives the most basic information 
for the experiments. Quantitative calculations of thermodynamic quantities 
such as pressure and energy density are indispensable. 
In particular, since the number density fluctuation should be large 
around the critical endpoint, 
the susceptibility of the quark number density, 
given by the second derivative of pressure with respect to $\mu_q$, 
is an important quantity. 
Several studies of the quark number susceptibility 
have been performed at $\mu_q=0$ \cite{Gott}. 
Moreover, measurements of pressure and energy density at 
$\mu_q \neq 0$ were done using the reweighting method \cite{FKS}, 
which allow the investigation of thermodynamic properties at 
non-zero baryon density. This approach, however, does not work 
for large $\mu_q$ and large lattice size due to the sign problem. 

Our strategy is the following. 
We compute the derivatives of physical quantities with respect to $\mu_q$ 
at $\mu_q=0$, and determine the Taylor expansion coefficients in terms of 
$\mu_q$\cite{TARO,us02}, in which the sign problem does not arise. 
Because the pressure is an even function of $\mu_q$, 
the $\mu_q^2$-term is leading and $\mu_q^4$-term is the next to leading,
and in fact only these two terms are non-zero 
in the high temperature (Stefan-Boltzmann) limit.
We compute the Taylor expansion coefficients up to fourth order. 
The fourth order term enables us to evaluate the $\mu_q$-dependence 
of the quark number susceptibility near $\mu_q=0$ . 
By estimating the change of the susceptibility, 
we discuss the possibility of the existence of the critical 
endpoint in the phase diagram of $T$ and $\mu_q$\cite{us03}. 

In Sec.~\ref{sec:transition}, 
we compute the Taylor expansion coefficients for the transition 
temperature $T_c$\cite{us02}. Some investigations of the phase structure on the 
$(T, \mu)$ plane have been done by the rewighting method \cite{Fod} and 
simulations with imaginary chemical potential \cite{dFP,DEL}. 
Because the procedure to identify the phase transition point is complex, 
it is difficult to apply the method which we use in Sec.~\ref{sec:pressure}. 
Our method for $T_c$ is the following. 
We use the reweighting method, but we perform a Taylor expansion for 
the quark determinant and the fermionic operators and neglect 
unnecessary higher order terms of $\mu_q$ for the calculation of 
the derivatives which we want. Then, 
in the region where the sign problem is not serious, we calculate 
the Taylor expansion coefficients. This method reduces computer time 
drastically in comparison with the original reweighting method and enables 
us to use rather large lattice size. 

Finally, we discuss the critical endpoint again. 
Since a Taylor expansion must break down at the singular point, 
it seems to be difficult to find the position of the critical endpoint by 
the Taylor expansion. We focus on a property of 3-flavor QCD at very 
small quark mass\cite{FK03}. 
The crossover phase transition becomes a first order phase transition 
at a critical quark mass also for $\mu_q=0$\cite{KLS}. 
Hence, by tracing out the position of the critical endpoint from that at 
$\mu_q=0$ to finite $\mu_q$ in the vicinity of $\mu_q=0$, 
we may be able to estimate the position of the critical endpoint 
for the real physical quark masses. 
We try it in Sec.~\ref{sec:critical}. 
Section \ref{sec:conclusions} presents our conclusions.

\section{Taylor expansion of pressure in terms of $\mu_q$}
\label{sec:pressure}

\begin{figure}[t]
\centerline{\epsfysize=2.2in\epsfbox{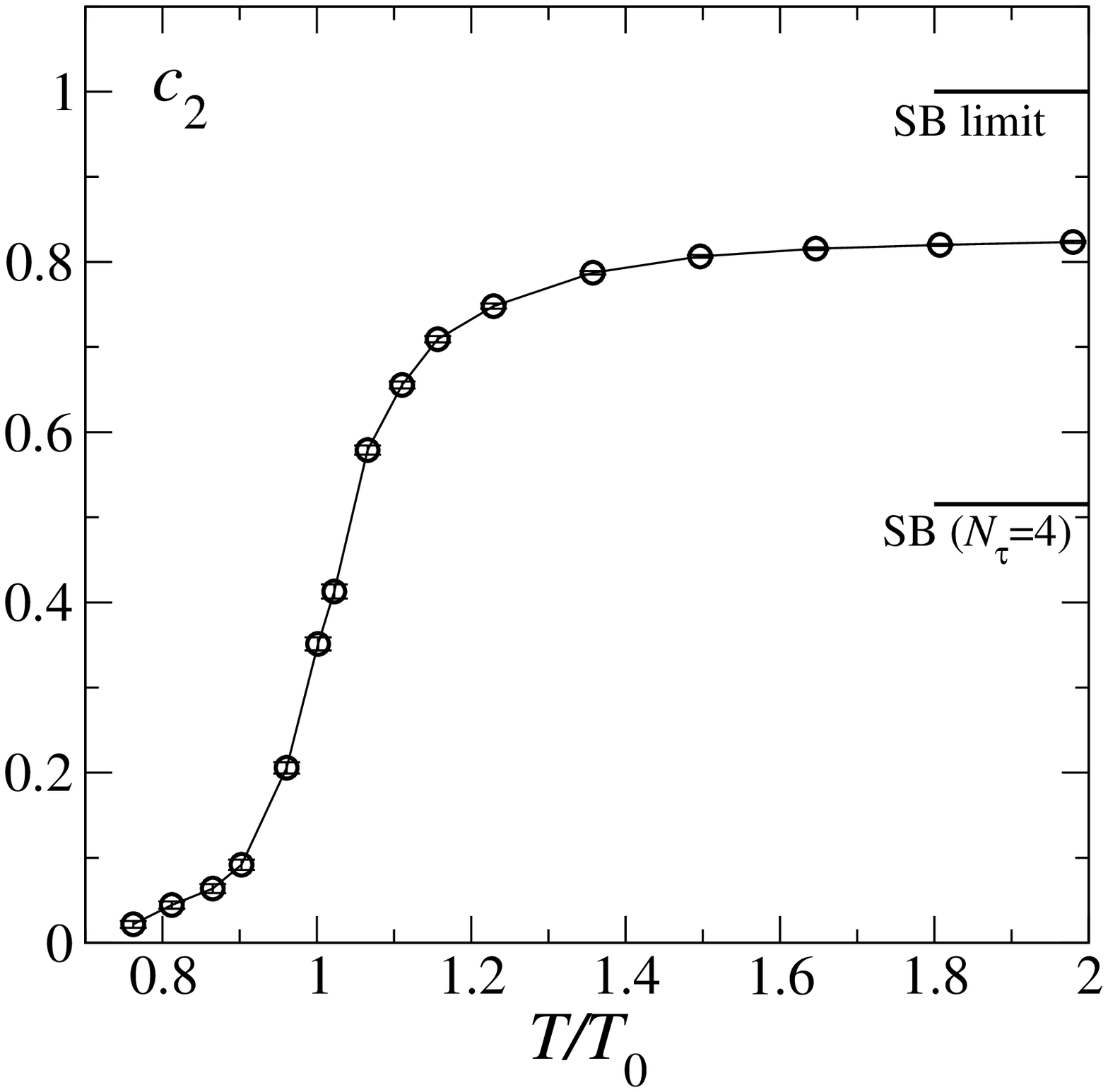} \hspace{5mm}
            \epsfysize=2.2in\epsfbox{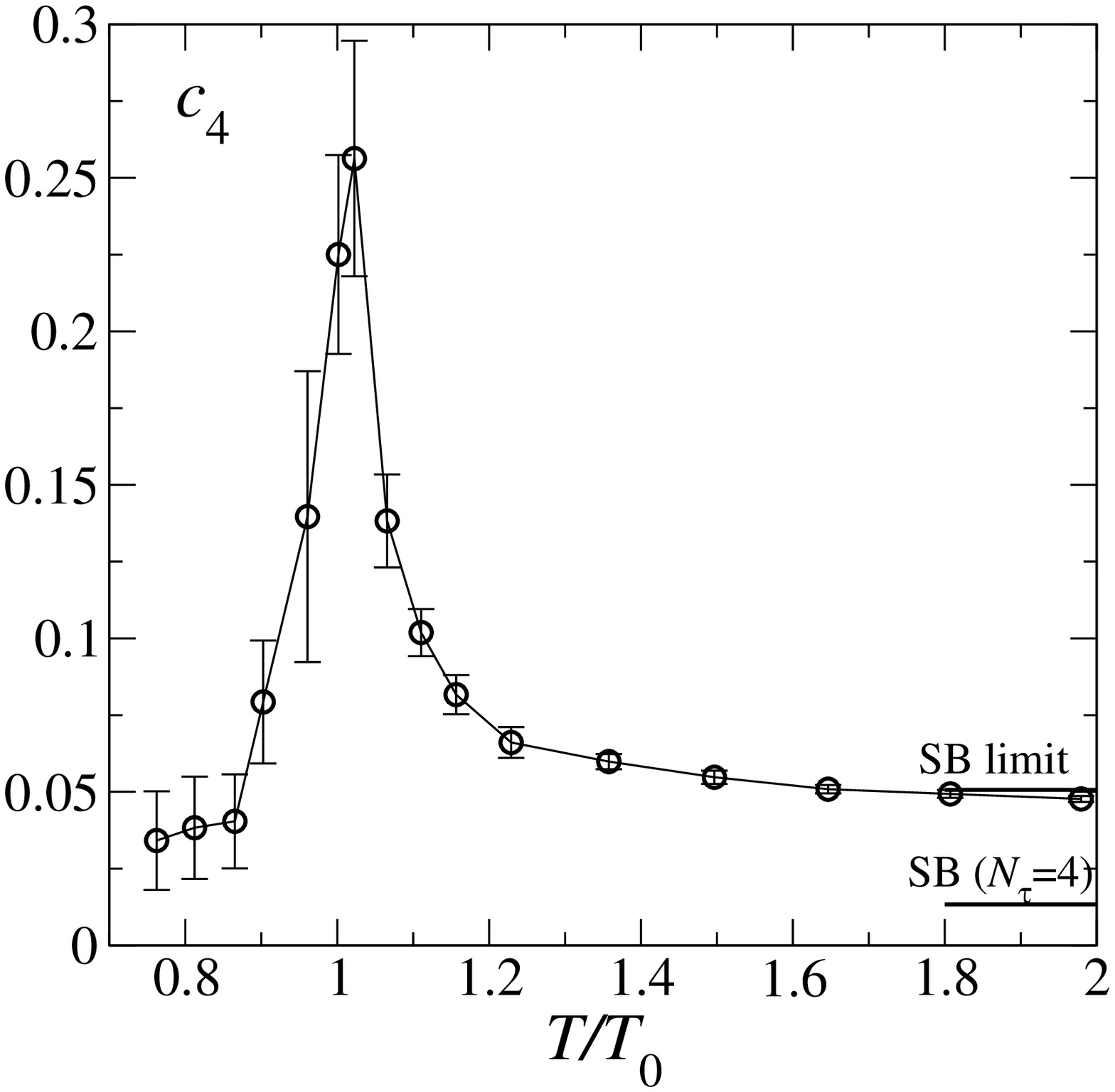}} 
\vspace{-2mm}
\caption{
Coefficients of Taylor expansion, $c_2$ (left) and $c_4$ (right).
$T_0$ is $T_c$ at $\mu_q=0$.
\label{fig:c2c4}}
\end{figure}

Pressure is given in terms of the grand partition function 
$Z(T, V, \mu_q)$ by
$p/T^4=(1/VT^3)\ln Z.$
However, the direct calculation of $\ln Z$ is difficult, hence 
most of the work done at $\mu_q=0$ for the calculation of pressure 
is done by using the integral method\cite{KLP,CPPACS}, 
where the first derivative of 
pressure is computed by simulations, and the pressure is obtained by 
integration along a suitable integral path. 
For $\mu_q \neq 0$, direct Monte Carlo simulation is not applicable; 
in this case we proceed by computing higher order derivatives of 
pressure with respect to $\mu_q/T$ at $\mu_q=0$, and then 
estimate $p(\mu_q)$ using a Taylor expansion, 
\begin{eqnarray}
\left. \frac{p}{T^4} \right|_{T,\mu_q} 
= \left. \frac{p}{T^4} \right|_{T,0} 
+ \sum_{n=1}^{\infty} c_n(T) \left( \frac{\mu_q}{T} \right)^n ,
\label{eq:taylorcont}
\end{eqnarray}
where $c_n=(1/n!) \partial^n(p/T^4)/\partial(\mu_q/T)^n |_{\mu_q=0}$.
Explicitly writing 
\begin{eqnarray}
c_2 = \frac{N_{\tau}}{2! N_{\sigma}^3} {\cal A}_2 , \hspace{10mm}
c_4 = \frac{1}{4! N_{\sigma}^3 N_{\tau}} ({\cal A}_4 -3 {\cal A}_2^2),
\end{eqnarray}
$
{\cal A}_2 =
\left\langle \frac{N_{\rm f}}{4} 
\frac{\partial^2 (\ln \det M)}{\partial \mu^2} \right\rangle 
+\left\langle \left( \frac{N_{\rm f}}{4} 
\frac{\partial (\ln \det M)}{\partial \mu} \right)^2 \right\rangle, 
$ \\ $
{\cal A}_4 =
\left\langle \frac{N_{\rm f}}{4} 
\frac{\partial^4 (\ln \det M)}{\partial \mu^4} \right\rangle 
+ 4 \left\langle \left( \frac{N_{\rm f}}{4} \right)^2 
\frac{\partial^3 (\ln \det M)}{\partial \mu^3}
\frac{\partial (\ln \det M)}{\partial \mu} \right\rangle 
+ 3 \left\langle \left( \frac{N_{\rm f}}{4} \right)^2 
\left( \frac{\partial^2 (\ln \det M)}{\partial \mu^2} 
\right)^2 \right\rangle 
\\ \hspace*{10mm}
+ 6 \left\langle \left( \frac{N_{\rm f}}{4} \right)^3 
\frac{\partial^2 (\ln \det M)}{\partial \mu^2} 
\left( \frac{\partial (\ln \det M)}{\partial \mu} \right)^2 
\right\rangle 
+ \left\langle \left( \frac{N_{\rm f}}{4} 
\frac{\partial (\ln \det M)}{\partial \mu} \right)^4 \right\rangle, 
$
\\
for staggered type quarks on an $N_{\sigma}^3 \times N_{\tau}$ lattice, 
and the coefficients for odd terms are zero.
$M$ is the quark matrix. 
$\mu$ is a chemical potential in lattice units, $\mu \equiv \mu_q a$. 
Here, we used a property that the odd derivatives of $\ln \det M$ are 
purely imaginary and the even derivatives are real\cite{us02}. 
These derivatives are computed by the random noise method, which 
saves computer time. 
Furthermore, we do not need simulations at $(T, \mu_q)=(0,0)$ 
for the subtraction to normalize the value of $p$, since 
the derivatives of $p|_{T=0,\mu_q=0}$ with respect to $\mu_q$ are, 
of course, zero. This also reduces computer time.

We compute the pressure up to $O(\mu_q^4)$ using 2-flavors of 
p4-improved staggered fermion\cite{HKS} at a bare quark mass $ma=0.1$ 
on a $16^3 \times 4$ lattice. 
Then the quark number density $n_q$ and quark number susceptibility 
$\chi_q$ are calculated up to $O(\mu_q^3)$ and $O(\mu_q^2)$, respectively. 
These are obtained by the derivatives of pressure; 
$n_q/T^3=\partial(p/T^4)/\partial(\mu_q/T)$,  
$\chi_q/T^2 = \partial^2 (p/T^4)/\partial (\mu_q/T)^2$, 
The details are given in Ref.~\citen{us03}. 
In Fig.~\ref{fig:c2c4}, we plot the data for $c_2$ (left) and 
$c_4$ (right). Both of them are very small at low temperature and 
approach the Stefan-Boltzmann (SB) limit in the high temperature 
limit, as expected. 
The remarkable point is a strong peak of $c_4$ around $T_c$. 

We can understand this peak via two arguments. 
One is a prediction from the hadron resonance gas model\cite{KRT}, 
which is an effective model of the free hadron gas in the low temperature 
phase. 
The model study predicts $c_4/c_2=0.75$ and our results are consistent 
with this prediction for $T < T_c$; in fact, as $T$ increases $c_4/c_2$ 
remains constant until $T \approx T_c$, 
whereupon it approaches the SB limit. 

The other point is from a discussion of the convergence radius of 
the Taylor expansion. 
We expect that the crossover at $\mu_q=0$ changes to a 
first order transition at a point $\mu_q/T_c \sim O(1)$\cite{Fod,FK03}. 
Then, the analysis by Taylor expansion 
must break down in that regime, i.e. the convergence radius should be 
smaller than the value of $\mu_q/T$ at the critical endpoint.
We define estimates for the convergence radius by 
$\rho_n \equiv \sqrt{|c_n / c_{n+2}|}$.
We compute $\rho_0$ and $\rho_2$ from $c_0 \equiv p/T^4(0)$, $c_2$ and $c_4$. 
It is found that both $\rho_0$ and $\rho_2$ are quite large at 
high temperature as expected from the SB limit,
$\rho_2^{SB} \simeq 2.01, \rho_4^{SB} \simeq 4.44.$ On the other hand, 
around $T_c$, these are $O(1)$, since $c_2$ and $c_4$ are of the same order, 
so that our results around $T_c$ suggest a singular point in the 
neighborhood of $\mu_q/T_c =1$.

Next, we calculate pressure and quark number susceptibility in a range 
of $0 \leq \mu_q/T \leq 1$ which is within the radius of convergence 
discussed above, 
Using the data of $c_2$ and $c_4$, $p/T^4$ and $\chi_q/T^2$ is obtained by
$\Delta(p/T^4) \equiv p(T,\mu_q)/T^4-p(T,0)/T^4=
c_2(\mu_q/T)^2+c_4(\mu_q/T)^4+O(\mu_q^6)$, 
and $\chi_q/T^2=2c_2+12c_4(\mu_q/T)^2+O(\mu_q^4)$. 

\begin{figure}[t]
\centerline{\epsfysize=2.1in\epsfbox{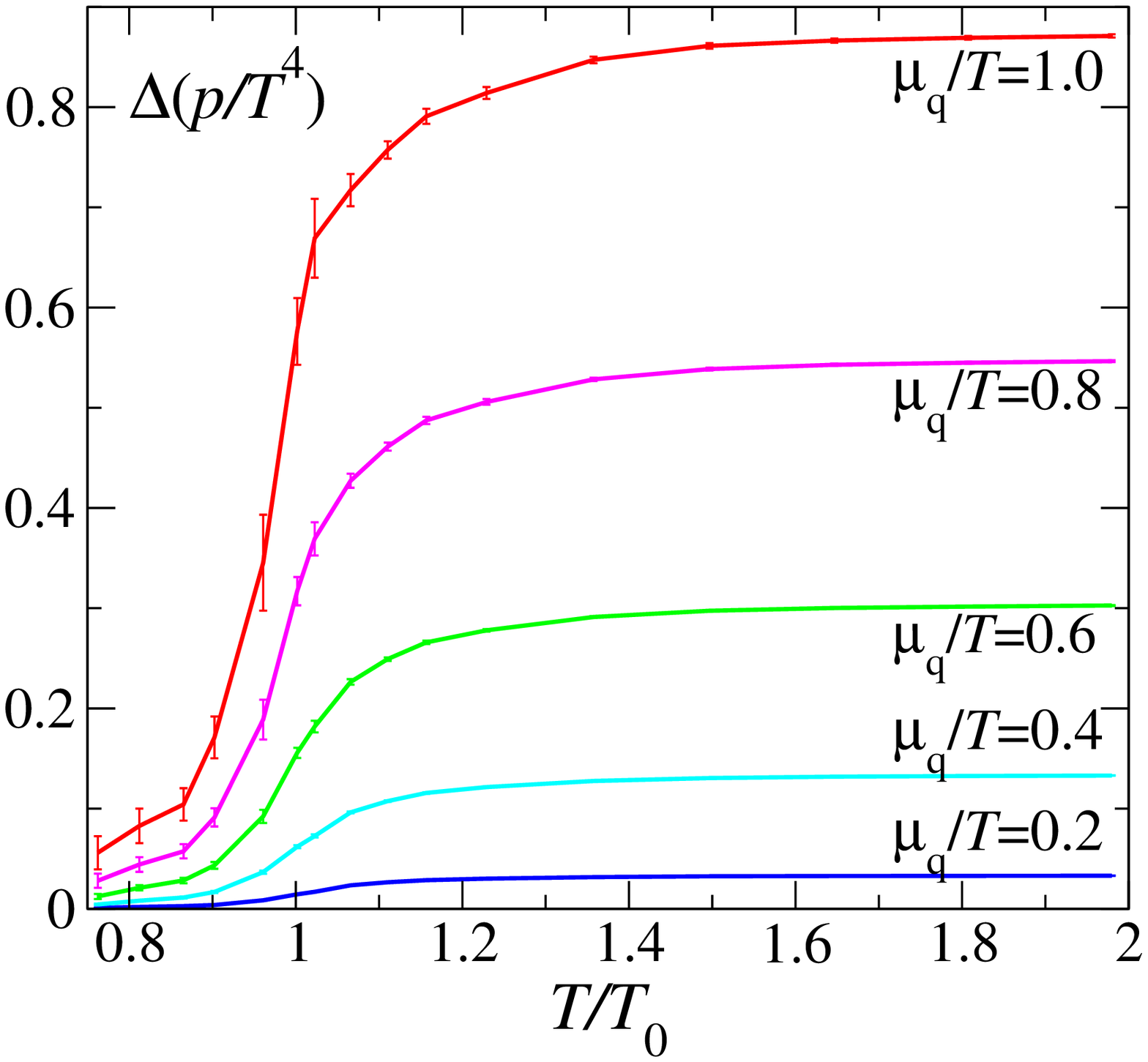} \hspace{7mm}
            \epsfysize=2.1in\epsfbox{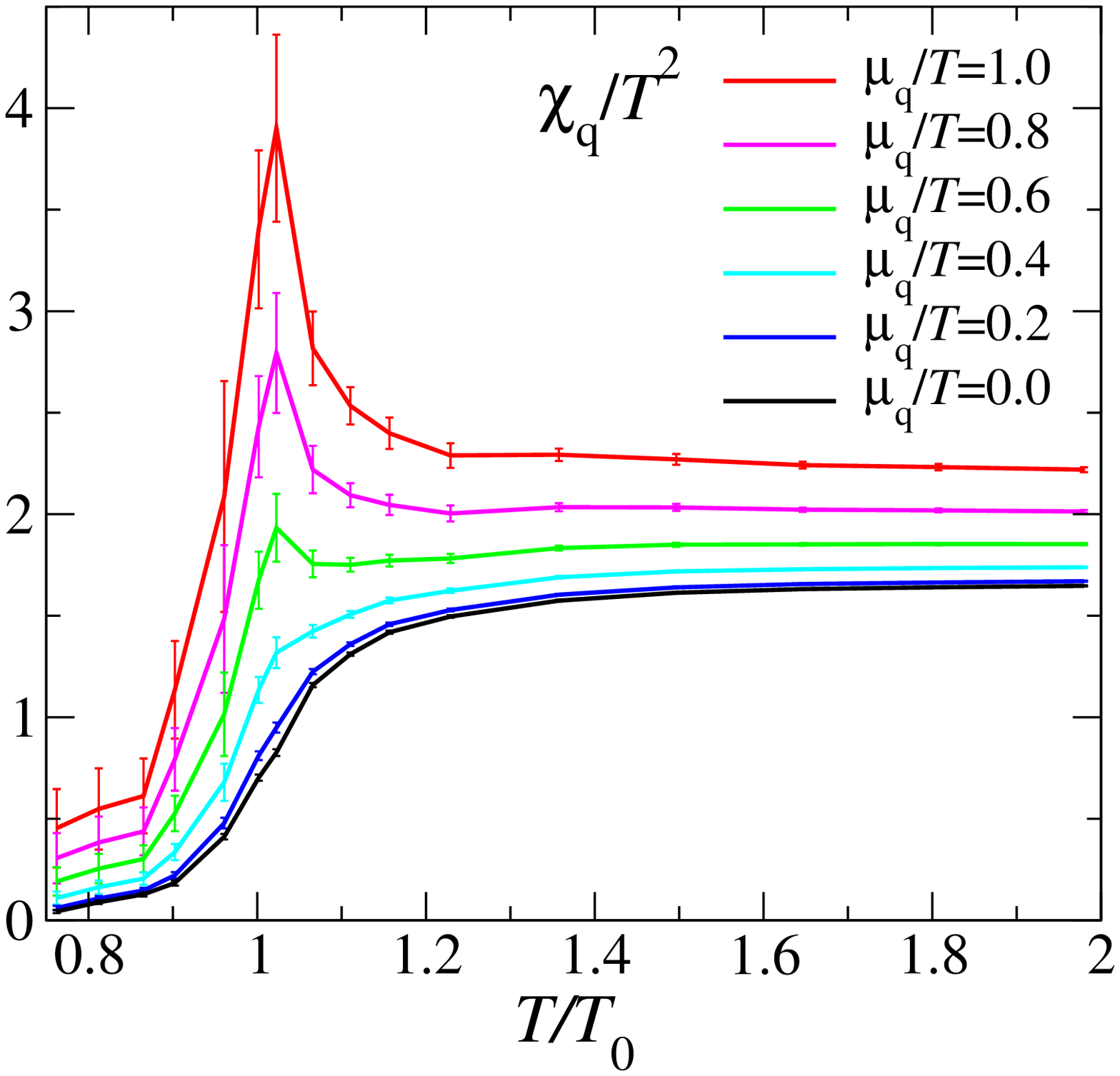}}   
\vspace{-2mm}
\caption{Difference of pressure from $\mu_q=0$ (left) and
Quark number susceptibility (right) as a function of $T$ 
for each fixed $\mu_q/T$.
$T_0$ is $T_c$ at $\mu_q=0$.
\label{fig:pres}}
\end{figure}

We draw $\Delta (p/T^4)$ for each fixed $\mu_q/T$ in Fig.~\ref{fig:pres} 
(left) and find that the difference from $p|_{\mu_q=0}$ is very small 
in the interesting regime for heavy-ion collisions, 
$\mu_q/T \approx 0.1$ (RHIC) and $\mu_q/T \approx 0.5$ (SPS), 
in comparison with the value at $\mu_q=0$, e.g. the SB value for 2-flavor QCD 
at $\mu_q=0$: $p^{SB}/T^4 \simeq 4.06$. The effect of non-zero quark density 
on pressure at $\mu_q/T=0.1$ is only $1\%$. 
Also, the result is qualitatively consistent with that 
of Ref.~\citen{FKS} obtained by the reweighting method. 

Figure \ref{fig:pres} (right) is the result for $\chi_q/T^2$ at fixed 
$\mu_q/T$. 
We find a pronounced peak for $m_q/T > 0.5$, whereas $\chi_q$ does 
not have a peak for $\mu_q=0$.
This suggests the presence of a critical endpoint in the $(T,\mu_q)$ plane.

This discussion can be easily extended to the charge fluctuation 
$\chi_C$.
The spike of $\chi_C$ at $T_c$ is weaker than that of 
$\chi_q$, which means the increase of the charge fluctuation is smaller 
than that of the number fluctuation as $\mu_q$ increases\cite{us03}.

\section{Phase transition line in the $(T, \mu_q)$ plane}
\label{sec:transition}

Next, we investigate the phase transition temperature as a function of 
$\mu_q$ by performing a Taylor expansion around $\mu_q=0$\cite{us02}. 
In principle, the method via a Taylor expansion is applied for any 
physical observable. 
However, as the order of the derivative becomes higher, the number of 
terms increases exponentially and there exist serious subtractions, 
so the method in the previous section does not suit a 
quantity which needs complex procedures for the measurement 
such as $T_c$. 

We consider the following identity, 
\begin{eqnarray}
\label{eq:rew}
\langle {\cal O} \rangle_{(\beta, \mu)} 
= \left\langle {\cal O} W \right\rangle_{(\beta_0,0)} /
\left\langle W \right\rangle_{(\beta_0,0)}, 
\end{eqnarray} 
where
$W = {\rm e}^{(N_{\rm f}/4) 
(\ln \det M(\mu) - \ln \det M(0))} {\rm e}^{-S_g(\beta)+S_g(\beta_0)}$ 
for staggered type quarks, $S_g$ is the gauge action, and $\beta=6/g^2$. 
An expectation value $\langle {\cal O} \rangle$ at $(\beta, \mu)$ is 
computed by a simulation at $(\beta_0, 0)$. 
This is a basic formula of the reweighting method, but we expand 
$\ln \det M(\mu)$ and any operator of a fermionic observable 
using a Taylor series, e.g. 
$\ln \det M(\mu) - \ln \det M(0) = \sum_{n=1}^{\infty} 
[\partial (\ln \det M)/\partial \mu] (\mu^n/n!)$.
This formula is valid in a region that the $\ln \det M(\mu)$ does not 
have any singular point for each configuration. 
We moreover neglect the terms higher than $O(\mu^n)$, 
then the resulting expectation value contains the error of 
the higher order of $\mu$ but the error does not affect the calculation 
of the derivatives for the lower order than the neglected terms.
In fact, if we expand into the form of the Taylor expansion from 
Eq.(\ref{eq:rew}), we obtain the correct coefficients of the Taylor expansion 
for $\langle {\cal O} \rangle$ up to $O(\mu^n)$,
and this expression is much simpler than the method discussed 
in the previous section. 
Here, we try to trace out the phase transition line for non-zero $\mu$ 
using this method.

However, if we want to use Eq.(\ref{eq:rew}), we must discuss a famous problem, 
called ``sign problem''. 
Because $\det M$ is complex at $\mu \neq 0$, if the complex phase 
fluctuates rapidly, the numerator and the denominator in RHS of 
Eq.(\ref{eq:rew}) becomes vanishingly small, and then we cannot obtain 
the reliable answer
for $\langle {\cal O} \rangle$. Of course, the phase fluctuation is zero at 
$\mu=0$, but it becomes larger as $\mu$ increases. 
However, in the context of the Taylor expansion, the applicable range of 
the reweighting method is rather easy to estimate by 
evaluating the complex phase fluctuation. 
For small $\mu \equiv N_{\tau}^{-1} (\mu_q/T)$, 
the complex phase can be written 
by the odd terms of the Taylor expansion of $\ln \det M$\cite{us02}. 
Denoting $\det M = |\det M| {\rm e}^{i \theta}$, 
\begin{eqnarray}
\label{eq:phase}
\theta = \frac{N_{\rm f}}{4} \sum_{n: {\rm odd}} {\rm Im} 
\left( \frac{\partial^n \ln \det M}{\partial \mu^n} \right) \mu^n.
\end{eqnarray}
The first term is
$(N_{\rm f}/4) {\rm Im} \, {\rm tr} [M^{-1} (\partial M/\partial \mu)] \mu$. 
From these equations, we find explicitly that the magnitude of 
$\theta$ is proportional to $\mu \equiv N_{\tau}^{-1} (\mu_q/T)$, 
$V \equiv N_{\sigma}^3$ and $N_{\rm f}$. Moreover this 
can be computed by the noise method. 
Roughly speaking, the sign problem happens when the fluctuation of 
$\theta$ becomes larger than $O(\pi/2)$. 
Thus the sign problem is not serious for small $\mu$ and small volume, 
but that region becomes narrower and narrower as the volume increases. 

We observed the complex phase fluctuation on a $16^3 \times 4$ 
lattice\cite{us02} and found that the applicable range is not so small 
for this lattice size. The applicable range is narrower than the convergence 
radius discussed in Sec.~\ref{sec:pressure}, but it covers 
the interesting regime for the heavy-ion collisions at RHIC. 
We also found a tendency for the phase fluctuation to increase as quark 
mass decreases, and that the fluctuation is small in the high temperature 
phase.

\begin{figure}[t]
\parbox{\halftext}{
\centerline{\epsfysize=2.2in\epsfbox{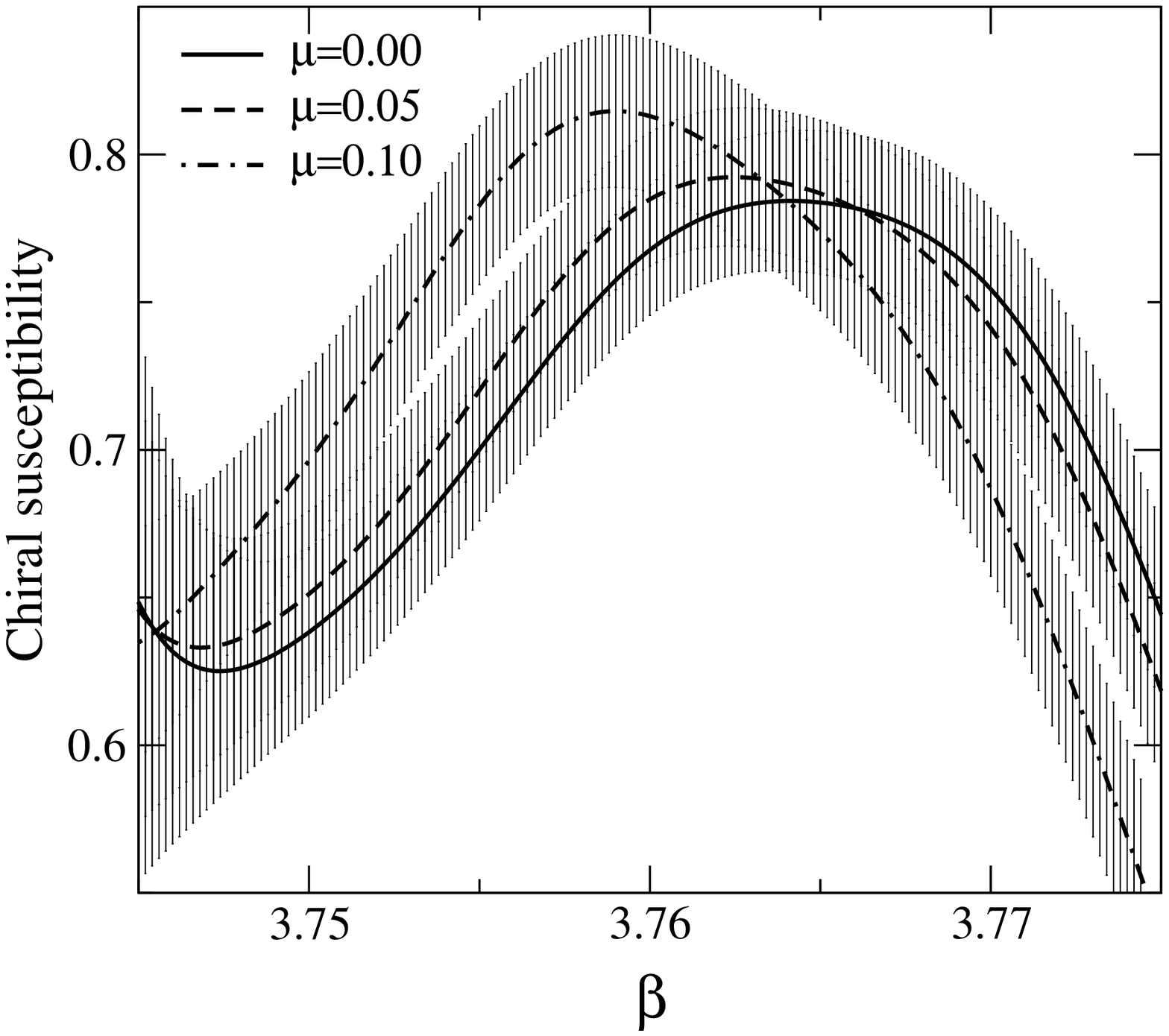}}
\vspace*{-2mm}
\caption{Chiral susceptibility.}
\label{fig:csu02}}
\hspace{3mm}
\parbox{\halftext}{
\centerline{\epsfysize=2.2in\epsfbox{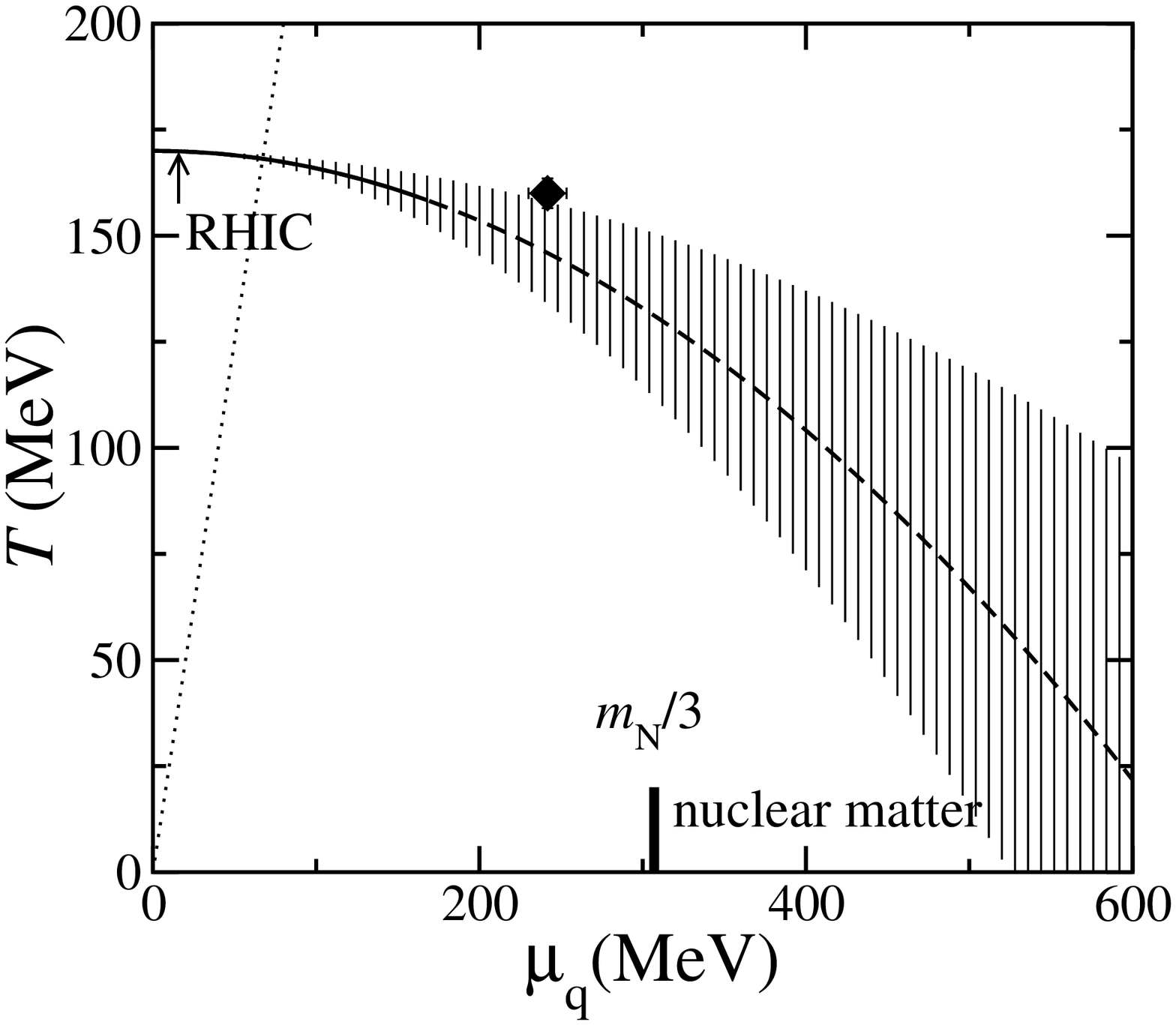}}   
\vspace*{-2mm}
\caption{Sketch of the phase diagram.}
\label{fig:cur}}
\end{figure}

In the region where the sign problem is not serious, 
we determine the phase transition line. 
Because the first derivative is expected to be zero from the symmetry 
under exchange of $\mu$ to $-\mu$, we calculate the second derivative 
of $T_c$ with respect to $\mu$. 
Simulations are performed around the phase transition point $\beta_c$ 
by using the p4-improved action at $ma=0.1$ and $0.2$, 
corresponding $m_{\pi}/m_{\rho} \approx 0.70$ 
and $0.85$, on a $16^3 \times 4$ lattice.
In Fig.~\ref{fig:csu02}, we plot chiral susceptibility 
as a function of $\beta$ for $\mu = 0, 0.05$ and $0.1$ at $ma=0.2$. 
This calculation contains errors of $O(\mu^4)$.   
From this figure, we find that the peak position of the susceptibility 
moves left as $\mu$ increases, which means that $\beta_c$ or $T_c$ 
decreases as $\mu$ increases.  
Assuming the peak position is $\beta_c$, 
we determine the second derivative of $\beta_c$. 
The truncation error of $O(\mu^4)$ does not affect to this calculation. 
Combining with the results for the Polyakov loop susceptibility, we obtain 
${\rm d}^2 \beta_c/{\rm d} \mu^2 \approx -1.1$ 
and the quark mass dependence of ${\rm d}^2 \beta_c/{\rm d} \mu^2$ for $ma=0.1$ 
and $0.2$ is not visible within the accuracy of our calculation. 
(However, we find large mass dependence for very small quark mass. 
See Sec.~\ref{sec:critical}.)

The second derivative of $T_c$ is given by 
\begin{eqnarray}
\frac{{\rm d}^2 T_c}{{\rm d} \mu_q^2} = -\frac{1}{N_{\tau}^2 T_c} 
\left. \frac{{\rm d}^2 \beta_c}{{\rm d} \mu^2} \right/ 
\left( a \frac{{\rm d} \beta}{{\rm d} a} \right),
\end{eqnarray}
The beta function, $a({\rm d} \beta / {\rm d} a)$, is obtained 
from the string tension data in Ref.~\citen{KLP}. 
We then find 
$(T_c/2)({\rm d}^2 T_c/{\rm d}\mu^2_q) = -0.07(3)$. 
We sketch the phase transition line from the curvature in Fig.~\ref{fig:cur}.
The diamond symbol is the critical point obtained by Ref.~\citen{Fod}. 
The dotted line is the upper bound of the fit range to determine the curvature.
At the relevant point for RHIC, this shift of $T_c$ is very small 
from that at $\mu=0$ 
and the result is roughly consistent with those obtained by 
the other groups\cite{Fod,dFP}.

Comparing the result of the equation of state, we also find that
the lines of constant pressure and energy density are roughly consistent 
with the phase transition line for small $\mu_q$, which suggest the change of 
pressure and energy density is small along the transition line. 
Moreover, by measuring the Polyakov loop, we can observe that the external 
quark current is screened by dynamical anti-quarks\cite{us02}.

\section{Critical endpoint for 3-flavor QCD}
\label{sec:critical}

Finally, we consider 3-flavor QCD. 
We perform simulations for $N_{\rm f}=3$ at $ma=0.1$ 
$(m_{\pi}/m_{\rho} \approx 0.68)$ on 
$16^3 \times 4$ lattices, and at $ma=0.005$ 
on $12^3 \times 4$ and $16^3 \times 4$ lattices, 
using the p4-improved action\cite{FK03}. 
The result of the curvature at $ma=0.1$ is 
$(T_c/2)({\rm d}^2 T_c/{\rm d}\mu^2_q) = -0.045(10)$, 
where we keep $\mu$ for strange quark equal to zero, and 
$a({\rm d} \beta / {\rm d} a)$ is obtained from 
the string tension data.
The difference from $N_{\rm f}=2$ is not large at $ma=0.1$. 
However, a significant difference is observed between $ma=0.005$ and $0.1$ 
for $N_{\rm f}=3$. Using a perturbative beta-function, 
we obtain 
$(T_c/2)({\rm d}^2 T_c/{\rm d}\mu^2_q) = -0.025(6)$, 
and $-0.114(46)$ for $m=0.1$ and $0.005$, respectively. 
Although the perturbative beta-function gives almost a factor of 2 smaller 
result than the non-perturbative analysis, 
this suggests that the curvature of the transition line becomes stronger 
as the quark mass decreases. 

The most interesting point for $N_{\rm f}=3$ is the existence 
of a critical quark mass $m_c$ on the $\mu_q=0$ axis, 
which separates a first order phase transition at small $m$ and 
crossover at large $m$. 
In order to identify $m_c$, 
we compute Binder cumulants constructed from the chiral condensate, 
$B_4= \langle (\delta \bar{\psi} \psi)^4 \rangle / 
\langle (\delta \bar{\psi} \psi)^2 \rangle^2$. 
It has been verified that the critical point belongs to 
the Ising universality class, 
i.e. the value of $B_4$ at $T_c$ for $m_c$ 
is the same with that of 3-dimensional Ising model \cite{KLS}.
Moreover, we expect such a critical point at $\mu_q \neq 0$ even 
in the region of large $m$, hence it is very interesting to 
investigate how the critical point moves as a function of $\mu_q$. 

We apply the Taylor expanded reweighting method, explained 
in Sec.~\ref{sec:transition}, with respect to quark mass, replacing 
$\mu$ by $\Delta m \equiv m-m_0$, where $m_0$ is a simulation point, and 
we investigate the quark mass dependence around the simulation point, 
$m_0=0.005$. The reweighting factor is computed up to $O[(\Delta m)^2]$. 
This analysis suggests a critical value of $m_c=0.0007(4)$ for $\mu_q=0$.
At this point, the Binder cumulant is consistent with the value of 
the Ising model, and also the peak height of the chiral susceptibility 
$\chi_{\bar{\psi}\psi}$ shows scaling behavior indicating a first order 
phase transition, i.e. $\chi_{\bar{\psi}\psi} \sim V$, 
for the volume size $V=12^3$ and $16^3$ below about this point. 
The corresponding value of pseudoscalar meson mass at the 
critical point is $m_{\pi}^{crit}=67(18) \ {\rm MeV}$. 
Hereafter, we use the value of $m_{\pi}$ at $T=0$ for setting a 
physical scale instead of the value of bare quark mass $m$. 

Next, we investigate the critical point for finite $\mu$ using 
the Taylor expanded reweighting method with respect to $\mu$, 
up to $O(\mu^2)$.
Since the sign problem is serious for the analysis on the 
$16^3 \times 4$ lattice, we analyze only the data on the 
$12^3 \times 4$ lattice. The data of $B_4$ suggests 
$\mu^{crit}=0.074(13)$ or $\mu_q^{crit}/T=0.296(52)$. 
An estimate for the transition temperature at corresponding 
$m_{\pi}$ is obtained from 
$T_c/\sqrt{\sigma}=0.40(1)+0.039(4)(m_{\pi}/\sqrt{\sigma})$\cite{KLP1}.
Then the critical value is $\mu_q^{crit}=52(10) \ {\rm MeV}$ at 
the simulation point $ma=0.005$. The corresponding $m_{\pi}$ is 
about $170 \ {\rm MeV}$. 

In the context of the Taylor expanded reweighting method for 
(2+1)-flavor QCD with equal up and down quark masses and 
a strange quark mass, $m_{ud}$ and $m_s$, 
the system is identical along the line of slope $-2$ in the 
$(m_{ud}, m_s)$ plane near the $m_{ud}=m_{s}$ line. Along this line, 
the first terms of reweighting factor in terms of $\Delta m$ 
for $m_{ud}$ and $m_s$ are cancelled, i.e. 
if $(m_s-m^{(3f)})/(m_{ud}-m^{(3f)})=-2$ is satisfied, the systems at 
$(m_{ud}, m_s)$ and $(m^{(3f)}, m^{(3f)})$ are the same for $m_{ud} \approx m_s$.
We use this property for the extrapolation to the physical point. 
Using the lowest order chiral perturbation theory relation, 
$m_K^2/m_{\pi}^2=(m_{ud}+m_s)/(2m_{ud})$, 
this line in the $(m_{ud}, m_s)$ plane is translated to the line of 
slope $-1/2$ in the $(m_{\pi}^2, m_K^2)$ plane. 
Hence, when $m_K^2=-m_{\pi}^2/2+3m_{\pi}^{(3f) 2}/2$, 
the physics at $(m_{\pi}^2, m_K^2)$ is roughly corresponding 
to that of 3-flavor QCD with $m_{\pi}^{(3f)}$.
For the physical value $m_{\pi}=140 \ {\rm MeV}$ and 
$m_{K} \approx 500 \ {\rm MeV}$, 
we obtain the corresponding pion mass for 3-flavor QCD, 
$m_{\pi}^{(3f)}=416 \ {\rm MeV}$. 
The extrapolation of the critical surface to this pion mass value 
is performed by a straight line in the 
$(m_{\pi}^2, \mu_q^2)$ plane using the two critical points 
which we found at $\mu_q=0$ and $\mu_q \neq 0$ for 3-flavor QCD. 
We obtain an estimate for the critical point at the physical 
quark mass, $\mu_q^{crit} \approx 140 \ {\rm MeV}$.
This value is smaller than the value of about $240 \ {\rm MeV}$ 
estimated in Ref.~\citen{Fod} for (2+1)-flavor QCD. 
Further study is clearly required to improve this analysis.

\section{Conclusions}
\label{sec:conclusions}

Taylor expansion coefficients were calculated up to $O(\mu^4)$ for pressure 
and up to $O(\mu^2)$ for phase transition temperature. 
We found that the differences of $T_c$ and pressure 
from those at $\mu_q=0$ are small for the interesting values of $\mu_q$ for 
the heavy-ion collisions at RHIC and SPS. 
The fluctuations in the quark number density 
increase in the vicinity of $T_c$ and 
the susceptibilities start to develop a pronounced peak 
as $\mu_q$ is increased. 
This suggests the presence of a critical endpoint in the 
$(T, \mu_q)$ plane.
Also, the critical endpoint for 3-flavor QCD near $\mu_q=0$ was investigated.
The extrapolation from the critical point at $\mu_q=0$ suggests 
that the critical endpoint exists 
around $\mu_q/T_c \sim 1$ for the physical quark mass.

\section*{Acknowledgements}

This work is supported by BMBF under grant No.06BI102, 
DFG under grant FOR 339/2-1 
and PPARC grant PPA/a/s/1999/00026.
SJH is supported by a PPARC Senior Research Fellowship.

\end{document}